\documentclass[twocolumn,twoside,dvipdfm,journal]{IEEEtran}
\usepackage{amssymb}
\usepackage{amsmath}
\usepackage{mathrsfs}
\usepackage{amsfonts}
\newtheorem{*theorem}{Theorem}
\usepackage{graphicx}
\usepackage{color}
\usepackage{cite}
\usepackage{amsmath}
\usepackage{extarrows}

\newcommand{\bw}{\mbox{\boldmath $w$}}
\newcommand{\bx}{\mbox{\boldmath $x$}}

\newcommand{\bQ}{\mbox{\boldmath $Q$}}

\newcommand{\bB}{\mbox{\boldmath $B$}}
\newcommand{\bG}{\mbox{\boldmath $G$}}
\newcommand{\bg}{\mbox{\boldmath $g$}}

\newcommand{\brho}{\mbox{\boldmath $\rho$}}

\newcommand{\HH}{\dagger}

\newtheorem{theorem}{Theorem}[section]

\newcommand{\bpsi}{\mbox{\boldmath $\psi$}}
\makeatletter

\newcommand{\Rmnum}[1]{\expandafter\@slowromancap\romannumeral #1@}
\makeatother

\begin{document}
\bibliographystyle{IEEEtran}
\title{\LARGE{Joint Power Splitting and Secure Beamforming Design in the Multiple Non-regenerative Wireless-powered Relay Networks}}

\author{Mingxiong~Zhao,~\IEEEmembership{Student Member,~IEEE,}
        Xiangfeng~Wang,
        and~Suili~Feng,~\IEEEmembership{Member,~IEEE}
\thanks{The work of M. Zhao and S. Feng was supported by the National Natural Science Foundation of China 61340035, the Science and Technology Program of Guangzhou 2014J4100246, and the Chinese Scholarship Council under Grant 201206150017. The work of X. Wang was supported by the Shanghai Youth Science and Technology Excellence Sail Plan 15YF1403400 and the Shanghai Science Fundation 15JC1401700. The authors would like to thank the supports of National Engineering Technology Research Center of Mobile Ultrasonic Detection.}
\thanks{M. Zhao and S. Feng are with the School of Electronics and Information Engineering, South China University of Technology, Guangzhou, China, 510640. Emails:~jimmyzmx@gmail.com,~fengsl@scut.edu.cn.}
\thanks{X. Wang is with Shanghai Key Laboratory of Trustworthy Computing, Software Engineering Institute, East China Normal University, Shanghai, China, 200062. Email:~xfwang@sei.ecnu.edu.cn.}}
\maketitle
\begin{abstract}
The physical-layer security issue in the multiple non-regenerative wireless-powered relay (WPR) networks is investigated in this letter, where the idle relay is treated as a potential eavesdropper. To guarantee secure communication, the destination-based artificial noise is sent to degrade the receptions of eavesdroppers, and it also becomes a new source of energy powering relays to forward the information with power splitting (PS) technique. We propose an efficient algorithm ground on \emph{block-wise penalty function method} to jointly optimize PS ratio and beamforming to maximize the secrecy rate. Despite the nonconvexity of the considered problem, the proposed algorithm is numerically efficient and is proved to converge to the local optimal solution. Simulation results demonstrate that the proposed algorithm outperforms the benchmark method.
\end{abstract}


\section{Introduction}

Simultaneous wireless information and power transfer (SWIPT) is a promising solution for prolonging the lifetime of energy-constrained wireless nodes \cite{ZhangHo}. Due to its efficient utilization of wireless spectrum to harvest energy and process information from the same wireless signal, SWIPT triggers a great deal of research interests in cooperative transmission \cite{EHrel,PSAS,ding2014power,krikidis2014simultaneous}. Two protocols, namely time switching (TS) and power splitting (PS), were proposed for energy harvesting relay with SWIPT in \cite{EHrel}. TS enabled relay utilize different time blocks to realize information processing (IP) and energy harvesting (EH), while relay proportionally split the received signal power to accomplish these two operations separately in PS protocol. In \cite{PSAS}, the authors investigated a multiple-antenna relay system with SWIPT where a ``harvest-and-forward" strategy was proposed to maximize the achievable rate. Meanwhile, the authors in \cite{ding2014power} investigated power allocation strategies and their impact on the multiple source-destination pair networks with an EH relay. And the performance of a large-scale network with multiple transmitter-receiver pairs with/without relaying was discussed in \cite{krikidis2014simultaneous}, where receivers employed PS technique to harvest energy.

Recently, with the rapidly developed SWIPT technologies, physical (PHY)-layer security  has attracted significant attention in SWIPT with application to wireless networks. As is mentioned in \cite{LiuGC13,NgGC13}, EH receivers are generally deployed sufficiently close to the transmitter than information receivers (IRs) to achieve higher signal power. This gives rise to a challenging PHY-layer security issue that EH receivers may easily eavesdrop the confidential information with better channels, rather than harvest energy as they are presumed to do. To resolve this issue, substantial research efforts have been dedicated to SWIPT with the consideration of PHY-layer security \cite{LiuGC13,NgGC13,Hong2015TVT,li2014robust}. The authors in \cite{LiuGC13} investigated secrecy wireless information and power transfer with joint information and energy beamforming design in multi-input single-output (MISO) system, while the dual use of artificial noise (AN) and energy signals for facilitating simultaneous secure communication and energy transfer with MISO beamforming was introduced in \cite{NgGC13}. The authors in \cite{Hong2015TVT,li2014robust} split the power at the transmitter to send confidential message to IR and an AN to interfere ER against eavesdropping. Moreover, the secrecy problem in cooperative networks with SWIPT was considered in \cite{li2014secure,EHjammer,xing2014harvest}. The authors in \cite{li2014secure} focused on secure relay beamforming for SWIPT in one-way relay systems with an external eavesdropper and EH receiver. Meanwhile, the authors in \cite{EHjammer,xing2014harvest} investigated a wireless-powered jammer system, where the jammer (e.g. idle EH receiver \cite{xing2014harvest}) could harvest energy from the wireless signal and use it to interfere with the external eavesdropper. However, these works \cite{li2014secure,EHjammer,xing2014harvest} associated with cooperative transmission did not consider that EH receivers may intercept the confidential information intended for IRs.

Motivated by the aforementioned problem, in this letter, we consider a multiple non-regenerative wireless-powered relay (WPR) networks, which is different from the scenario in \cite{PSAS} with only one WPR. In this networks, WPRs are supposed to harvest energy to assist the source-destination transmission when active. However, when the relay is idle, it may become a potential eavesdropper in the network, and maximize its decoding signal-to-noise ratio (SNR), rather than split power for cooperative transmission. To improve the security of the source-destination transmission, the destination-based AN is applied to our scenario to degrade the receptions of idle relays, which also becomes a new source of energy powering WPRs in the second-hop transmission. Distinguished from the exhaustive search for PS ratio in \cite{PSAS}, we propose an algorithm based on \emph{block-wise penalty function method} to find the local optimal PS ratio and optimize beamforming for each WPR to maximize the secrecy rate with much less computational time. Despite the nonconvexity of the considered problem, the proposed algorithm is numerically efficient and is proved to converge to the local optimal solution. Numerical experiments manifest the superior performance of the proposed scheme as compared with the benchmark method.

\emph{Notation:} We adopt the notation of using boldface for column vectors (lower case), and matrices (upper case). The hermitian transpose is denoted by the symbol $(\cdot)^\HH$. For a complex scalar $x$, its complex conjugate is denoted by $x^*$. $E[\cdot]$ and $\cal{CN(\cdot)}$ denote the statistical expectation and complex Gaussian distributions, respectively. For vector $\bx$, diag($\bx$) means putting $\bx$ on the main diagonal.

\section{System model}
Consider a two-hop parallel relay networks where a source $\mathcal{S}$, a destination $\mathcal{D}$ and a set of relay nodes ($r_i, i=1,\cdots,N$) are all equipped with a single antenna. Each $r_i$ can assist the source-destination transmission when active. In this paper, we assume that the active relay $r_i$ can not only process the received signal, but also harvest energy from it to help $\mathcal{S}$ forward the information resorting to PS technique. However, when the relay is idle, it may become a potential eavesdropper in the network, and maximize its decoding SNR, rather than split the power for cooperative transmission. What's worse, if the idle relay is closer to $\mathcal{S}$ or other active relays, it can easily eavesdrop the confidential information with better channels.

To guarantee secure communication, at the time of confidential signal transmission, $\mathcal{D}$ concurrently sends AN to the area of relay nodes to degrade the channel conditions of idle relays. At the second hop, $r_i$ amplifies and forwards the received signal to $\mathcal{D}$ depending on its harvested energy. In this circumstances, AN can not only be used to degrade the receptions of eavesdroppers, but also become a new source of energy powering the information delivering from relays to $\mathcal{D}$ \cite{NgGC13}. This scenario is potentially applicative for sensor networks where a sink node with conventional supply of power wants to send its data to sensors far away from it. So the sink node asks for the help of some intermediary sensor nodes. However, the intermediary sensor is not willing to help the sink node due to the limited battery capacity. Hence, SWIPT is employed to encourage those intermediary sensors to use the harvested energy to participate in cooperative communication \cite{xiong2015wireless}. Nevertheless, in this application, some idle sensors may prefer to eavesdrop the legitimate information, rather than help the cooperative transmission, which incurs serious challenges to the PHY-layer security.

Denote the channel responses from $\mathcal{S}$ to $r_i$, from $\mathcal{D}$ to the active $r_i$ and from $r_i$ to $\mathcal{D}$ by $h_{si}\in \mathbb{C}$, $h_{di}\in \mathbb{C}$ and $h_{id}\in \mathbb{C},\ \forall i$. At the first hop, $\mathcal{S}$ sends the legitimate signal $x_s$ to relay nodes, meanwhile, $\mathcal{D}$ transmits AN $x_d$ to confound the potential eavesdroppers in the area of relay nodes. The received signal at relay $r_i$ can be expressed as 
\begin{equation}
\label{received signal}
y_i=h_{si}x_s+h_{di}x_d+n_i,
\end{equation}
where $n_i \in \mathbb{C}$ represents the additive white Gaussian noise (AWGN) associated with relay node $r_i$ following the distribution ${\cal{CN}}(0,\sigma^2_i)$, meanwhile, $x_s$ and $x_d$ are the transmitted symbols of $\mathcal{S}$ and $\mathcal{D}$ with the conventional energy supplies, such as $E[x_sx_s^*]=P_s$ and $E[x_dx_d^*]=P_d$. Based on PS scheme, $\rho_i$ denotes the ratio of power split for EH, and the left power for IP is $(1-\rho_i)$ at $r_i$. Hence, the parts of received signal for EH and IP at $r_i$ are written as follows, respectively
\begin{eqnarray}
y_i^{\text{EH}}&=&\sqrt{\rho_i}(h_{si}x_s+h_{di}x_d+n_i),\label{EH}\\
y_i^{\text{IP}}&=&\sqrt{1-\rho_i}(h_{si}x_s+h_{di}x_d+n_i)+n_{ci}\label{IP},
\end{eqnarray}
where $n_{ci} \in \mathbb{C}$ is the additive noise introduced by signal conversion from the passband to baseband with the distribution ${\cal{CN}}(0,\sigma^2_{ci})$ at $r_i$. The harvested energy of $r_i$ is given by
\begin{equation}
\label{harvested energy}
P_i^{\text{EH}}=\eta\rho_i\left(P_s|h_{si}|^2+P_d|h_{di}|^2+\sigma_i^2\right),
\end{equation}
where $\eta\in (0,1)$ denotes the energy conversion efficiency from signal power to circuit power.

In order to get more insight of this relay networks, we assume that there is only one idle relay ($r_{N+1}$) eavesdropping the legitimate information in this letter, and denote it as the potential eavesdropper ($\mathcal{E}$). The channels from $\mathcal{S}$ to $\mathcal{E}$, from $\mathcal{D}$ to $\mathcal{E}$ and from $r_i$ to $\mathcal{E}$ are given by $h_{se}$, $h_{de}$ and $h_{ie},\forall i$. Therefore, the received signal-to-interference-and-noise ratio (SINR) of $\mathcal{E}$ at the first hop can be written as $\text{SINR}_{e,1}=\frac{P_s|h_{se}|^2}{P_d|h_{de}|^2+\sigma_e^2}$, and $\sigma_e^2$ is the AWGN at $\mathcal{E}$.

We denote $w_i$ as the scalar beamformer at relay $r_i$, and utilize it to deal with the received signal at the corresponding relay node. Assuming perfect self-interference cancellation, the received signal of $\mathcal{D}$ at the second hop can be given by
\begin{eqnarray}
y_d\!\!\!\!&=&\!\!\!\!\sum\nolimits_{i=1}^N \sqrt{1-\rho_i}h_{id}w_ih_{si}x_s+\nonumber\\
\!\!\!\!&&\!\!\!\!\sum\nolimits_{i=1}^N \sqrt{1-\rho_i}h_{id}w_in_i+\sum\nolimits_{i=1}^N h_{id}w_in_{ci}+n_d,
\end{eqnarray}
where $n_d$ is the AGWN introduced by the receiver antenna at $\mathcal{D}$ with the distribution ${\cal{CN}}(0,\sigma_d^2)$. For notational convenience, we make $\sigma_i^2=\sigma_{ci}^2=\sigma_e^2=\sigma_d^2=\sigma^2,\ \forall i$ in the following parts. Making use of matrix-vector expressions, the achievable rate of $\mathcal{D}$ can be written as
\begin{equation}
\label{rate of dest}
R_d(\bw,\brho)\!\!=\!\!\text{log}_2\!\left(\!1\!+\!\frac{\bw^\HH\bB_{\rho}\bG_1\bB_{\rho}\bw}{\bw^\HH\bB_{\rho}\bQ_1\bB_{\rho}\bw+\bw^\HH\bQ_1\bw+\sigma^2}\!\right),
\end{equation}
where $\bG_1=P_s\bg_1\bg_1^\HH$, and $\bg_1=[h_{1d}h_{s1};\cdots;h_{Nd}h_{sN}]\in \mathbb{C}^{N\times 1}$. $\bw=[w_1;\cdots;w_N]\in \mathbb{C}^{N\times 1}$, $\brho=[\rho_1;\cdots;\rho_N]\in\mathbb{C}^{N\times 1}$,
$\bB_{\rho}=\text{diag}\left(\sqrt{1-\rho_1},\cdots,\sqrt{1-\rho_N}\right)$, and
$\bQ_1=\text{diag}\left(|h_{1d}|^2,\cdots,|h_{Nd}|^2\right)\sigma^2.$

To maximize the eavesdropping rate, $\mathcal{E}$ combines the two received signals in both stages using maximal ratio combining (MRC), and its achievable rate is given by
\begin{eqnarray}
\label{rate of eavesdropper}
\!\!\!\!\!\!\!\!R_e(\bw,\brho)\!\!\!\!\!\!\!\!\!\!\!\!&&=\!\text{log}_2\left(1+\frac{P_s|h_{se}|^2}{P_d|h_{de}|^2+\sigma^2}+\right.\nonumber\\
&&\!\!\!\!\left.\frac{\bw^\HH\bB_{\rho}\bG_2\bB_{\rho}\bw}{\bw^\HH\bB_{\rho}\bG_3\bB_{\rho}\bw\!\!+\!\!\bw^\HH\bB_{\rho}\bQ_2\bB_{\rho}\bw\!\!+\!\!\bw^\HH\bQ_2\bw\!\!+\!\!\sigma^2}\!\right),
\end{eqnarray}
where $\bG_2=P_s\bg_2\bg_2^\HH$ and $\bg_2=[h_{1e}h_{s1};\cdots;h_{Ne}h_{sN}]\in\mathbb{C}^{N\times 1}$; $\bG_3=P_d\bg_3\bg_3^\HH$, $\bg_3=[h_{1e}h_{d1};\cdots;h_{Ne}h_{dN}]\in\mathbb{C}^{N\times 1}$, and $\bQ_2=\text{diag}\left(|h_{1e}|^2,\cdots,|h_{Ne}|^2\right)\sigma^2$.

\section{Problem Formulation For Secrecy Rate Maximization}
In this section, we investigate the joint PS and beamforming design to maximize the secrecy rate $R_{sr}$, which is given by
\begin{equation}
R_{sr}(\bw,\brho) = \frac{1}{2} \left[R_d(\bw,\brho)-R_e(\bw,\brho)\right]^+,
\end{equation}where $[a]^+=\max(0,a)$, and $\frac{1}{2}$ is from the fact that the source-destination transmission takes place in two time slots.

For the considered scenario in this paper, the optimization problem can be written as
\begin{eqnarray}\label{optimization problem1}
\max_{\bw,\brho} && R_{sr}(\bw,\brho)\\
\text{s.t.} && |w_i|^2 \left[(1 - \rho_i)P_s|h_{si}|^2 + (1 - \rho_i)P_d |h_{di}|^2 + \right. \nonumber\\
&&\left. (2 - \rho_i)\sigma^2\right] \leq P_i^{\text{EH}}, \rho_i \in [0,1],\forall i = 1,\cdots,N,\nonumber
\end{eqnarray}
where the transmit power of $r_i$ is constrained by the harvested energy at $r_i$, $\forall i$. It is noted that the considered optimization problem \eqref{optimization problem1} is non-convex, and the combined inequality constraints further increase the difficulty to globally solve \eqref{optimization problem1} with respect to the two coupled blocks of variables $(\bw,\brho)$. In order to handle the nonconvexity and inequality constraints, we introduce the penalty function method \cite{courant1943variational} into our problem. Before that manipulation, formulation \eqref{optimization problem1} should be equivalently transformed into the following problem with the newly introduced slack variable $\bpsi$,
\begin{eqnarray}\label{optimization problem with barrier}
\max_{\bw,\brho,\bpsi} && R_{sr}(\bw,\brho)\\
\text{s.t.} && |w_i|^2 \left[(1 - \rho_i)P_s|h_{si}|^2 + (1 - \rho_i)P_d |h_{di}|^2 + \right. \nonumber\\
&&\left. (2 - \rho_i)\sigma^2\right] + \psi_i - P_i^{\text{EH}} = 0, \rho_i \in [0,1],\nonumber \\
&&  \psi_i \ge 0, \forall i = 1,\cdots,N.\nonumber
\end{eqnarray}
Further let us define the following penalty function
\begin{eqnarray}\label{penalty_function}
f(\bw,\brho,\bpsi,\lambda) \!\!\!\!\!\!&=&\!\!\!\!\!\! R_{sr}(\bw,\brho)\!\!-\!\! \lambda\sum\nolimits_{i=1}^N \left| |w_i|^2\left[(1-\rho_i)P_s|h_{si}|^2 \right.\right.\nonumber\\
+ ( 1 - \rho_i ) P_d \!\!\!\!\!\!\!\!\!&&\!\!\!\!\!\!\!\!\! \left.\left.|h_{di}|^2 + (2 - \rho_i) \sigma^2 \right] + \psi_i - P_i^{\text{EH}} \right|^2,
\end{eqnarray}where the second term represents the penalty term constituted by a total of $N$ constraints in \eqref{optimization problem with barrier}, $\lambda$ denotes the penalty parameter. Notice that problem \eqref{optimization problem with barrier} is equivalently rewritten as the maximization problem of \eqref{penalty_function} constrained by the feasible set of \eqref{optimization problem with barrier}, in which the penalty term equals zero. If the obtained solutions are not feasible, which violate the constraints of \eqref{optimization problem with barrier}, the penalty term may increase much with the quadratic growth, resulting in lower value of \eqref{penalty_function}. The penalty function is introduced to balance the secrecy-rate performance against the penalty term through the setup of $\lambda$. Therefore, the key idea of penalty function method is to increase the penalty parameter $\lambda$ progressively in order to obtain an optimal feasible solution of \eqref{optimization problem with barrier}.

In consideration of the problem structure, we further introduce \emph{Block Coordinate Gradient Descent} (BCGD) Method \cite{tseng2009coordinate} into our problem, which can be utilized to handle the large-scale multiple-block problem by computing and unifying each block variable through sequential updating. And the target of each update in our proposed algorithm is to increase the value of \eqref{penalty_function} by reducing the penalty term or enhancing $R_{sr}$. With the help of Armijo stepsize rule (refer to Table \Rmnum{1}), the monotonic increment of \eqref{penalty_function} is guaranteed in each update. When the stopping criteria is met according to Table \Rmnum{1}, the constraints of \eqref{optimization problem with barrier} are all satisfied, making the penalty term equal zero, and the local optimal solutions are obtained.

The block-wise penalty function method is a combination of penalty function method and BCGD method, the details of which are given in Table \Rmnum{1}, where $\epsilon$ is the tolerant error, $\bw_{t(i)}^k$, $\brho_{t(i)}^k$ and $\bpsi_{t(i)}^k$ are the temporary vectors with latest updated information instead of $\bw^k$, $\brho^k$ and $\bpsi^k$,
\begin{eqnarray}\label{notation_temp}
\bw_{t(i)}^k &=& (w_1^{k+1};\cdots;w_{i-1}^{k+1};w_i^k;w_{i+1}^k;\cdots;w_N^k),\nonumber\\
\brho_{t(i)}^k &=& (\rho_1^{k+1};\cdots;\rho_{i-1}^{k+1};\rho_i^k;\rho_{i+1}^k;\cdots;\rho_N^k),\nonumber\\
\bpsi_{t(i)}^k &=& (\psi_1^{k+1};\cdots;\psi_{i-1}^{k+1};\psi_i^k;\psi_{i+1}^k;\cdots;\psi_N^k).\nonumber
\end{eqnarray}

\begin{table}
\caption{The Block-wise Penalty Function Method}
\begin{center}
\begin{tabular}{|l|}
\hline
Initialization: Generate feasible block variables $(\bw^0,\brho^0,\bpsi^0)$ \\
\quad and penalty parameter $\lambda^0$. \\
\text{S1}: \textbf{For} $k=1,\cdots,K$, do S2--S3 until converge\\
\text{S2}: \textbf{For} $i=1,\cdots,N$, do\\
\quad (1) $w_i^{k+1}=w_i^{k} + \alpha_{w_i} d_{w_i}(\bw_{t(i)}^k,\brho^k,\bpsi^k,\lambda^k)$, $\forall i$; \\
\quad (2) $\rho_i^{k+1} = \rho_i^{k} + \beta_{\rho_i} d_{\rho_i}(\bw^{k+1},\brho_{t(i)}^k,\bpsi^k,\lambda^k)$, $\forall \rho_i\in [0,1]$;\\
\quad (3) $\psi_i^{k+1} = \psi_i^{k} + \gamma_{\psi_i} d_{\psi_i}(\bw^{k+1},\brho^{k+1},\bpsi_{t(i)}^k,\lambda^k)$, $\forall \psi_i \geq 0$. \\
\quad \quad Note that $\alpha_{w_i}$, $\beta_{\rho_i}$ and $\gamma_{\psi_i}$ are the step sizes in each update,\\
\quad and can be obtained by\\
\quad\quad \emph{Armijo stepsize rule}: For the considered maximization problem,\\
\quad choose $0<\alpha^{\text{init}}\leq 1$ and $\hat{\sigma},\beta \in (0,1)$. Let $\alpha^{\text{max}}$ be the largest\\
\quad element in $\{{{\alpha^{\text{init}}\beta^m}\}}_{m=0,1,\cdots}$ satisfying\\
\qquad\quad $f(x^{k+1})-f(x^k)\geq \hat{\sigma}\alpha^{\text{max}}\nabla f(x^k)'\hat{d}^k,$\\
\quad where $x^k$ can be replaced by $w_i^k$, $\rho_i^k$ and $\psi_i^k,\ \forall i$, and $\nabla f(x^k)$\\
\quad turns to be the partial derivative of $f$ with respect to $x^k$, i.e.,\\
\quad $\frac{\partial f}{\partial x}(x^k)$. And $\hat{d}^k$ is the ascent direction for our problem.\\
\quad \quad The ascent direction $d_{w_i}(\bw_{t(i)}^k,\brho^k,\bpsi^k,\lambda^k)$ is set as follows\\
\quad \qquad $d_{w_i}(\bw_{t(i)}^k,\brho^k,\bpsi^k,\lambda^k) = -\frac{\partial f}{\partial w_i}(\bw_{t(i)}^k,\brho^k,\bpsi^k,\lambda^k).$\\
\quad Similarly, the other two $d_{\rho_i}(\bw^{k+1},\brho_{t(i)}^k,\bpsi^k,\lambda^k)$ and \\
\quad $d_{\psi_i}(\bw^{k+1},\brho^{k+1},\bpsi_{t(i)}^k,\lambda^k)$ can be achieved.\\
\text{S3}: Stopping Criteria (Sequential Convergence)\\
\quad\ \ \textbf{If} $ \min\left\{\frac{\|\bw^{k+1}-\bw^k\|^2}{\|\bw^{k+1}\|^2}, \frac{\|\brho^{k+1}-\brho^k\|^2}{\|\brho^{k+1}\|^2}, \frac{\|\bpsi^{k+1}-\bpsi^k\|^2}{\|\bpsi^{k+1}\|^2}\right.$ \\
\quad\quad\quad $\left. \sum_{i=1}^N \left| |w_i^{k+1}|^2\left[(1\!-\!\rho_i^{k+1})P_s|h_{si}|^2 \!+\!(1\!-\!\rho_i^{k+1} ) P_d |h_{di}|^2 \right.\right.\right.$\\
\quad\quad\quad $\left. \left. \left. +(2 - \rho_i^{k+1}) \sigma^2 \right] +\psi_i^{k+1} - P_{i,{k+1}}^{\text{EH}} \right|^2 \right\} \leq \epsilon $, \\
\qquad\quad Stop and return $(\bw^{k+1},\brho^{k+1},\bpsi^{k+1})$;\\
\quad\ \ \textbf{Else} Set $\lambda^{k+1}=c\lambda^{k} (c\ge 1)$, $k:=k+1$ and go to \text{S1}.\\
\quad\ \ \textbf{End}\\
  \hline
\end{tabular}
\end{center}
\end{table}
Before presenting the convergence result, we first assume the feasible set is nonempty, which means that there is at least one feasible point of \eqref{optimization problem with barrier}. Next we have the following result:
\begin{theorem}
Any limit point of the sequence $\{\bw^k, \brho^k, \bpsi^k\}$ generated by the proposed block-wise penalty function method is a Karush--Kuhn--Tucker (KKT) point of problem \eqref{optimization problem with barrier}.
\end{theorem}

{\bf{Proof}}. First let $(\bw^*, \brho^*, \bpsi^*)$ be any limit point of the sequence $\{\bw^k, \brho^k, \bpsi^k\}$. Due to the property of Armijo stepsize rule, we have the monotonic increment of $f(\bw^k,\brho^k,\bpsi^k,\lambda^k)$, and obtain the following inequality
\begin{eqnarray}\label{equation1}
\!\!\!\!f(\bw^0,\brho^0,\bpsi^0,\lambda^0)\!\!\!\!\!&<&\!\!\!\!\!f(\bw^k,\brho^k,\bpsi^k,\lambda^k)\nonumber\\
\!\!\!\!\!&=&\!\!\!\!\!R_{sr}(\bw^k,\brho^k\!)-\lambda^k\mu(\bw^k,\brho^k,\bpsi^k).
\end{eqnarray}
Rearranging \eqref{equation1}, we have
\begin{equation}
\lambda^k\mu(\bw^k,\brho^k,\bpsi^k)\!\!<\!R_{sr}(\bw^k,\brho^k\!)-\!f(\!\bw^0,\brho^0,\bpsi^0,\lambda^0\!)\!<\!\infty,\!\!\!
\end{equation}
where $\mu(\bw^k,\brho^k,\bpsi^k)$ represents the second term without $\lambda$ in \eqref{penalty_function}. Considering the boundedness of $R_{sr}$ and the property of the limit point $(\bw^*,\brho^*,\bpsi^*)$, we have $\lambda^\infty\mu(\bw^*,\brho^*,\bpsi^*)<\infty$. Because $\lambda^\infty\to \infty$, we can deduce $\mu(\bw^*,\brho^*,\bpsi^*)=0$ and $(\bw^*,\brho^*,\bpsi^*)$ is a feasible point of \eqref{optimization problem with barrier}. If $(\bw^*,\brho^*,\bpsi^*)$ is not a KKT point, there should be another feasible point $(\tilde{\bw}, \tilde{\brho}, \tilde{\bpsi})$ making
\begin{equation}\label{noequal}
R_{sr} (\tilde{\bw}, \tilde{\brho}) > R_{sr} (\bw^*, \brho^*).
\end{equation}
Further, let us consider the subsequence $\{k_i\}$, where $\lim\limits_{i\to\infty}(\bw^{k_i},\brho^{k_i},\bpsi^{k_i})=(\bw^*,\brho^*,\bpsi^*)$, so that
$$
\lim_{i\rightarrow \infty} f(\!\bw^{k_i},\!\brho^{k_i},\!\bpsi^{k_i},\!\lambda\!)\!=\!\!f(\!\bw^*,\brho^*,\bpsi^*,\lambda\!)\!\mathop {\rm{\!\!=\!\!}}\limits^{(a)}\!R_{sr} (\!\bw^*,\!\brho^*\!),\forall \lambda,\nonumber
$$
where (a) is from the fact that the penalty term of \eqref{penalty_function} equals zero at point $(\bw^*,\brho^*,\bpsi^*)$
With a view to \eqref{noequal}, for large enough $k_i$, we have
$$
f(\!\tilde{\bw}, \tilde{\brho}, \tilde{\bpsi},\lambda^{k_i}\!) \!\!\mathop {\rm{\!\!=\!\!}}\limits^{(b)}\! R_{sr} (\!\tilde{\bw}, \tilde{\brho}\!)\!\!>\!\!R_{sr}(\!\bw^*,\brho^*\!)\!\!\geq\!\!f(\!\bw^{k_i},\!\brho^{k_i},\! \bpsi^{k_i},\!\lambda^{k_i}\!),\nonumber
$$
which is a contradiction to the $k_i$-th iteration with respect to $\lambda^{k_i}$, where (b) is due to the fact that $(\tilde{\bw}, \tilde{\brho}, \tilde{\bpsi})$ is a feasible point of \eqref{optimization problem with barrier} making the second term of \eqref{penalty_function} equal zero.$\hfill\square$

\section{Simulation Results}\label{simulation results}
\begin{figure}
  \centering
   \includegraphics[height=40mm, origin=br]{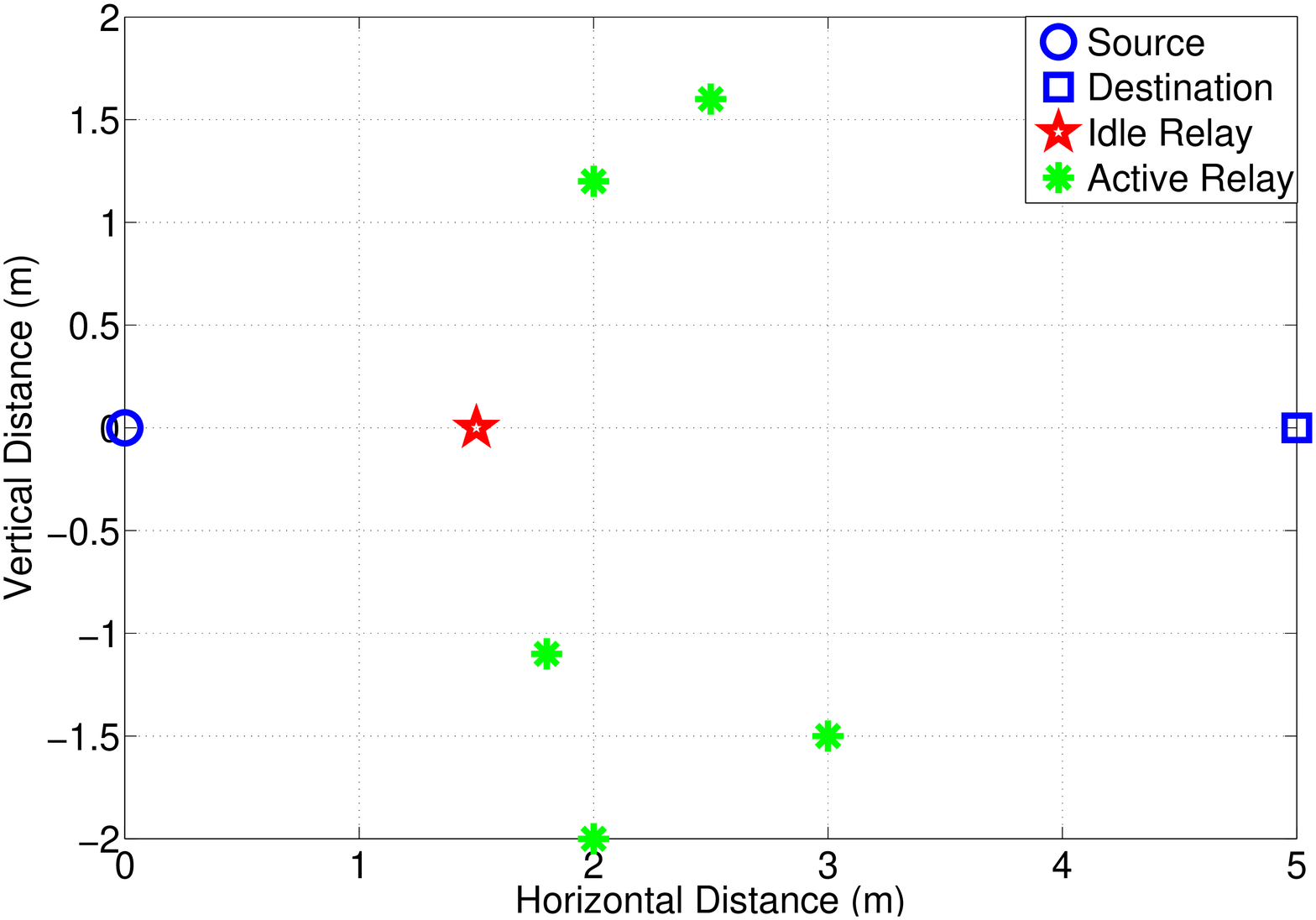}\\
   \caption{Simulation Scenario.}\label{1}
   \includegraphics[height=40mm, origin=br]{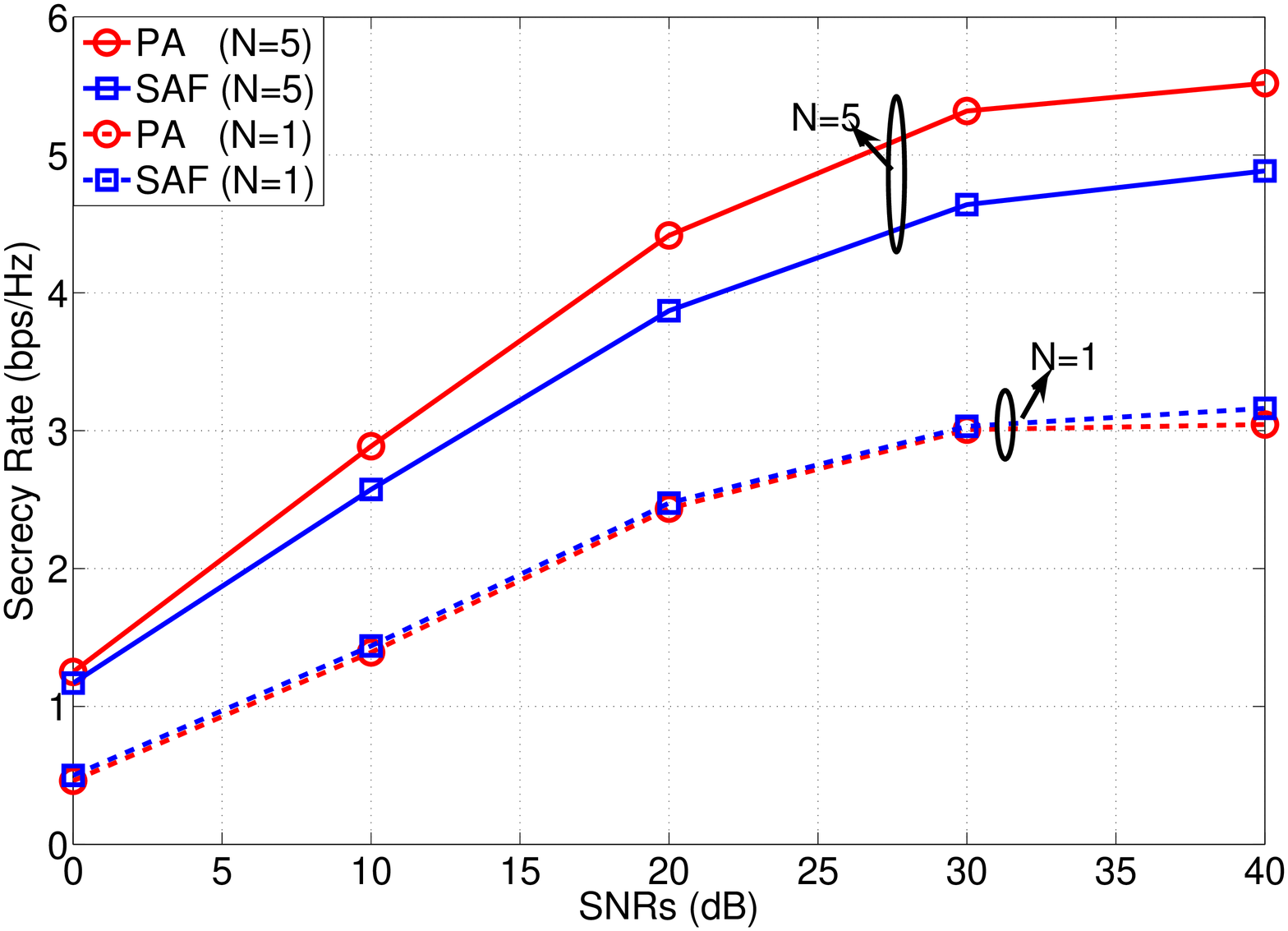}\\
   \caption{Secrecy rate versus $\text{SNR}_s$, $P_d=40\text{dBm}$, $\text{SNR}_s=\frac{P_s}{\sigma^2}$.}\label{2}
   \includegraphics[height=40mm, origin=br]{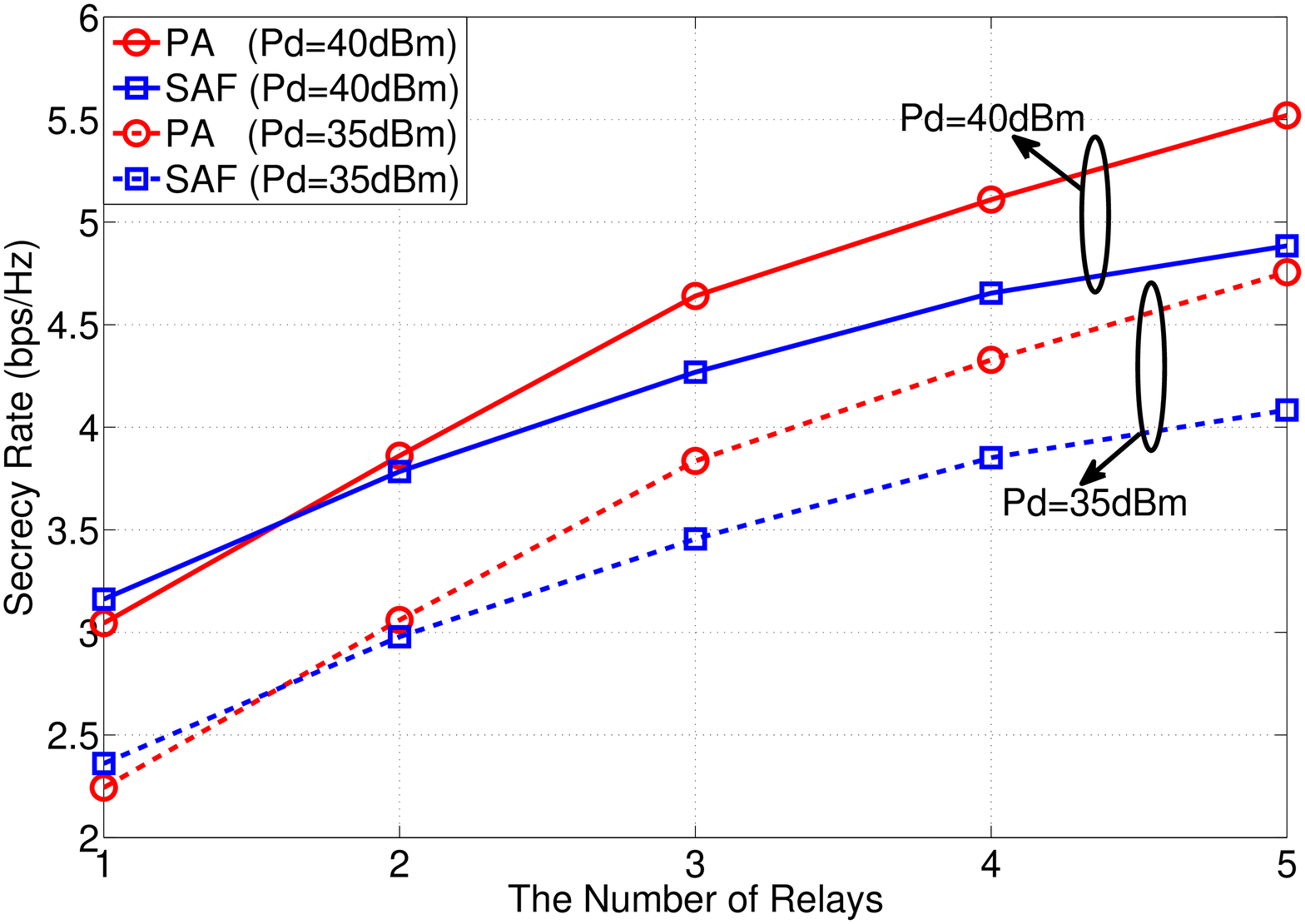}\\
   \caption{Secrecy rate versus different number of relays, $P_s=40\text{dBm}$.}\label{3}
   \includegraphics[height=40mm, origin=br]{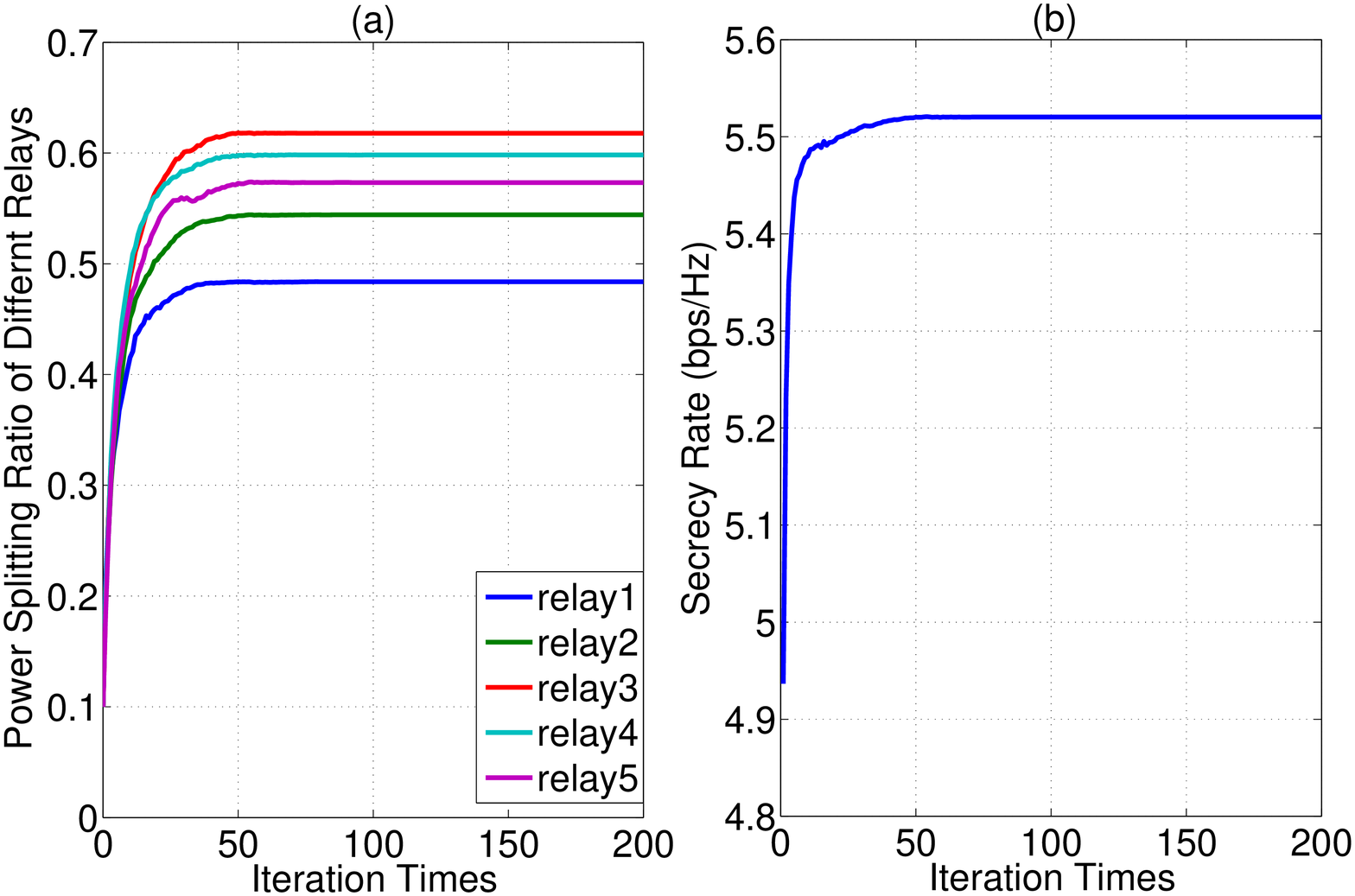}\\
   \caption{(a) Convergence results of $\rho$ for each relay;\ (b) Secrecy rate versus iteration times achieved by PA. $N=5$, $P_s=P_d=40\text{dBm}$ for (a) and (b).}\label{4}
\end{figure}

In this section, we investigate the performance of the proposed algorithm (PA) via numerical simulations. The path loss model for the energy harvesting relay channel is Rayleigh distributed and denoted by $|\beta|^2d^{-2}$, where $|\beta|$ and $d$ represent the short-term channel fading and the distance between two nodes. $|\beta|^2$ follows the exponential distribution with unit mean\cite{PSAS}. The position of each node is given in Fig.\ref{1}. We set the noise power as $\sigma^2=-40\text{dBm}$, the energy conversion efficiency $\eta=0.5$, the penalty parameter $\lambda=35$, the constant $c=1.2$ and the initial PS ratio $\rho_i=0.1,\ \forall i$ for all simulations results, which are averaged over 100 channel realizations.

A benchmark method is set as a comparison, named as simple amplify-and-forward (SAF) scheme, where we perform exhaustive search to obtain the power splitting ratio for each relay, and apply a simple amplified scalar $\theta_i$ with the consideration of power constraint at the relay, where $\theta_i=\sqrt{P_i^{\text{EH}}/\left[(1-\rho_i)(P_s|h_{si}|^2+P_d|h_{di}|^2+\sigma_i^2)+\sigma_{ci}^2\right]},\forall i$.

The comparisons of secrecy rate performance achieved by PA and SAF are evaluated in Fig.\ref{2} with fixed $P_d=40\text{dBm}$. For the special case $N=1$, it is observed that SAF achieves a better performance than PA. The reason is that SAF can get the global optimal PS ratio by exhaustive search in this case, where PA just obtains a local optimal one, but the performance of PA approaches to be optimal in the whole region of $\text{SNR}_s$, which indicates that PA is accurate and effective. When $N=5$, PA outperforms SAF in the whole region of $\text{SNR}_s$.

In Fig.\ref{3}, we show the performances yielded by different number of relays and by different $P_d$ resorting to PA and SAF with fixed $P_s=40\text{dBm}$, respectively. When the power of AN is enhanced, both PA and SAF enjoy a better secrecy rate due to the weakened reception of eavesdropper. Moreover, it is noted that as the number of relays increases, more WPRs are participating in cooperative transmission yielding higher performances, meanwhile, PA achieves the maximum secrecy rate and the increasing performance gap with $\!N\!$ is notably seen.

Fig.\ref{4} is presented to demonstrate the convergence result of PA, where we give the convergence result of PS ratio and the secrecy rate versus iteration times. Notice that the PS ratio $\rho$ of each relay will finally converge to its own local optimal solution as the iteration times increases.

\section{Conclusions}
In this paper, we investigate the secrecy rate maximization problem in multiple non-regenerative wireless-powered relays networks based on power splitting scheme, which has not been studied yet in the literature. We propose an algorithm resorting to block-wise penalty function method to jointly optimize power splitting and beamforming for all active relays, which outperforms the benchmark. And we also prove that the proposed algorithm can converge to a local optimal solution.

\bibliographystyle{IEEEbib}
\bibliography{reference}

\end{document}